\documentclass[pdflatex,sn-mathphys]{sn-jnl}

\jyear{2023}%

\theoremstyle{thmstyleone}%
\theoremstyle{thmstyletwo}%

\theoremstyle{thmstylethree}%

\def\P3M{P$^3$M}

\def\mooon{$\ddot{\mu}$}
\def\mooons{$\ddot{\mu}$s}

\begin{document}

\title[COWSHED II]{Galaxy Evolution in \mooon\ based Cosmologies}


\author*[1]{\fnm{Will} J. \sur{Roper}}\email{W.Roper@sussex.ac.uk}

\author[1]{\fnm{Stephen} M. \sur{Wilkins}}\email{S.Wilkins@sussex.ac.uk}

\author[1]{\fnm{Stephen} \sur{Riggs}}\email{S.Riggs@sussex.ac.uk}

\author[1]{\fnm{Jessica} \sur{Pilling}}\email{J.Pilling@sussex.ac.uk}

\author[2]{\fnm{Aswin} P. \sur{Vijayan}}\email{apavi@space.dtu.dk}

\author[3]{\fnm{Dimitrios} \sur{Irodotou}}\email{Dimitrios.Irodotou@helsinki.fi}

\author[1]{\fnm{Violetta} \sur{Korbina}}\email{V.Korbina@sussex.ac.uk}

\author[1]{\fnm{Jussi} \sur{Kuusisto}}\email{J.Kuusisto@sussex.ac.uk}

\affil*[1]{\orgdiv{Astronomy Centre}, \orgname{University of Sussex}, \orgaddress{\street{Falmer}, \city{Brighton}, \postcode{BN1 9QH}, \country{UK}}}

\affil[2]{\orgdiv{DTU-Space}, \orgname{Technical University of Denmark}, \orgaddress{\street{Elektrovej 327}, \city{Kongens Lyngby}, \postcode{2800}, \country{Denmark}}}

\affil[3]{\orgdiv{Department of Physics}, \orgname{University of Helsinki}, \orgaddress{\street{Gustaf Hällströmin katu 2}, \city{Helsinki}, \postcode{FI-00014}, \country{Finland}}}


\abstract{We present the first study of galaxy evolution in \mooon\ based cosmologies. We find that recent JWST observations of massive galaxies at extremely high redshifts are consistent with such a cosmology. However, the low redshift Universe is entirely divergent from the \mooon\ cosmic star formation rate density. We thus propose that our Universe was at one point dominated by a Primordial Bovine Herd (PBH) which later decayed producing dark energy. Note that we do not detail the mechanisms by which this decay process takes place. Despite its vanishingly small probability for existence, a \mooon\ based cosmological model marries the disparate findings in the high and low redshift Universe.}




\maketitle

\section{Introduction}\label{sec:intro}

In an infinite multiverse, anything can happen. Admittedly, for extremely unlikely events this is only true with a vanishingly small probability. However, no matter how small that probability is, it is nonetheless non-zero. In that spirit, we move the COWS all tHE way Down (COWSHED) project \citep{Roper2022} onto new pastures with another seminal paper, this time investigating galaxy formation and evolution in a universe dominated by a Primordial Bovine Herd (PBH) formed of \mooons\ (pronounced ``moo-on/s'').

\subsection{A \mooon\ based universe}

In a \mooon\ based cosmology, the density field established during inflation is perfectly arranged such that baryons are capable of condensing at recombination, not just into atoms but into fully realised cows, the titular \mooon\ (macroscopic) ``particles''. 

\begin{figure}[H]
    \centering
    \includegraphics[width=0.8\textwidth]{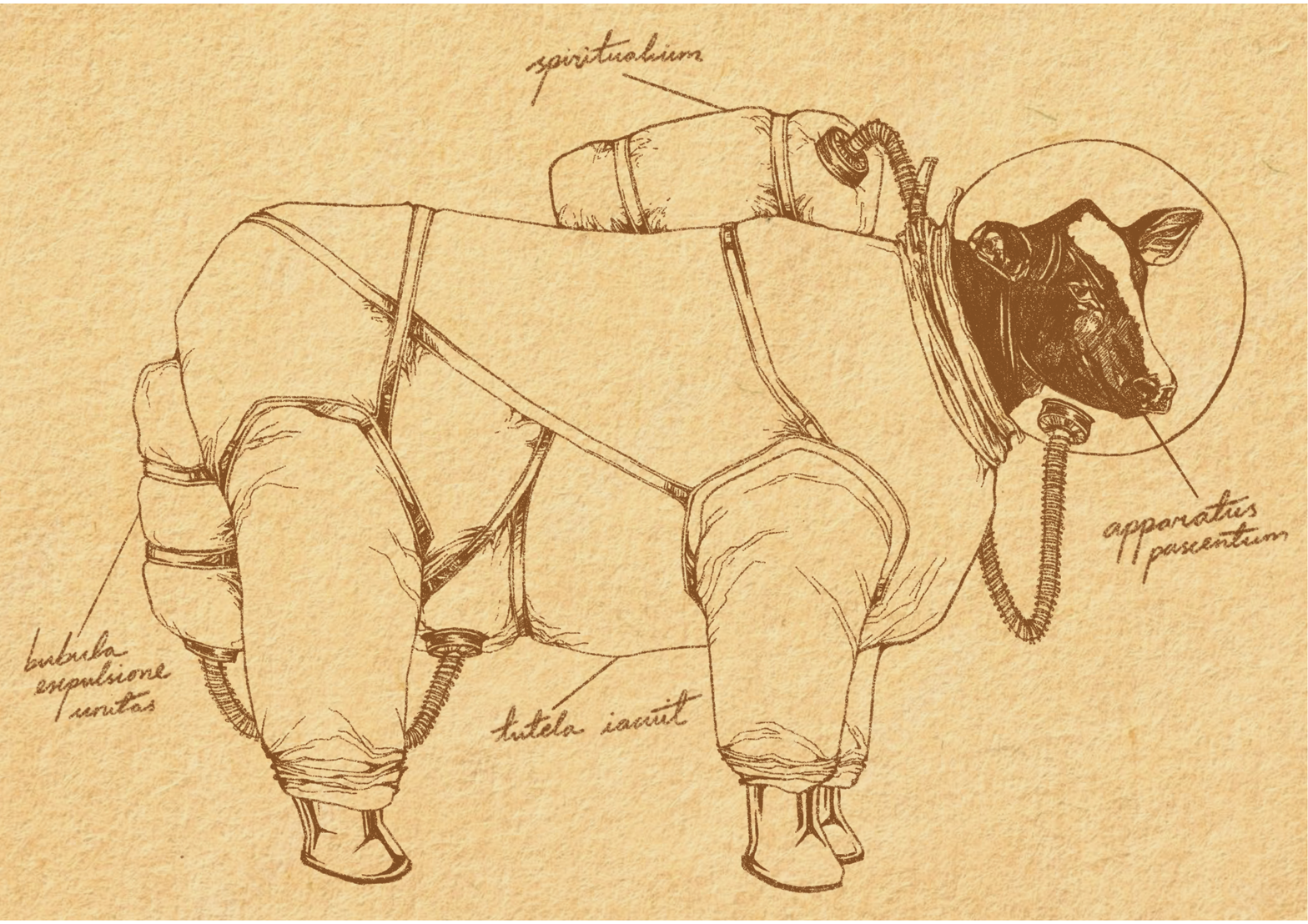}
    \caption{An artist's impression of a \mooon. To aid the reader we have labelled some important \mooon ian features of note, giving them their scientific names: spiritualium, apparatus pascentium, bubula expulsione unitas, tutela iacuit. Artist credit: Jessica Pilling.}
    \label{fig:mooon}
\end{figure}

Of course, there are numerous requirements to make such a Universe viable. If our intrepid bovines do not condense with the appropriate protection from the ultra-high redshift environment they are done for. We present an artist's impression of a successful \mooon\ in Figure \ref{fig:mooon}. Similarly, the bovines must have a sufficient source of sustenance and be able to procreate. Given the only conceivable baryonic ``food'' available is bovine (or bovine product) in nature we leave the specifics of primordial bovine diet and \mooon\ reproduction to the reader's imagination. Evidently, there are numerous failed \mooon\ based universes in the multiverse making the existence of a successful \mooon\ dominated universe \textbf{almost} impossible. But non-zero probability is nonetheless non-zero. We leave the exact mechanisms behind the success of a PBH for future works. We note that regardless of the survival mechanisms employed in the early universe, by $z\sim127$ the temperature of the Cosmic Microwave Background is $T\sim300$ K making the Universe hospitable for \mooons\ to eke out a comfortable existence \cite{Loeb2014}. 

This paper is structured as follows. In Section \ref{sec:sim} we detail the simulations used in this work. In Section \ref{sec:cow_subgrid} we detail the subgrid model used to describe the evolution of \mooons\ and the galaxies\footnote{Quite entertainingly things have come full circle. The origins of the word galaxy can be traced back to the Greek \textit{g\'{a}la} meaning milk. Sometimes research just feels predestined.}
that form from them. In Section \ref{sec:results} we detail the evolution of galaxies in a \mooon\ dominated universe.

To aid comparison with our own Universe we adopt a Planck year 1 cosmology ($\Omega_{0} = 0.307$, $\Omega_\Lambda = 0.693$, $h = 0.6777$, \cite{planck_collaboration_2014}) and a Chabrier stellar initial mass function (IMF) \citep{chabrier_galactic_2003}. However, we note that a \mooon\ based universe is not isolated to a cosmology like that in our own Universe. A \mooon\ based cosmology introduces a new cosmological parameter, namely the bovine fraction $f_{\rm bov}$ which encodes the fraction of baryons comprised of \mooons. In this work, we mainly consider $f_{\rm bov}=1$ but also briefly consider a universe with $f_{\rm bov}=0.5$.

We highlight that this work also has value for non-\mooon\ based cosmologies. Since galaxy evolution will proceed in much the same way once established, the findings of this work can be interpreted as an extreme case of early enrichment in the gas from which the first stars form. 

\section{The Simulations}
\label{sec:sim}

In this work, we use a small suite of simulations: two (12.5 cMpc)$^3$ periodic volumes with different bovine fractions ($f_{\rm bov}=[0.5, 1]$) and a single (50 cMpc)$^3$ periodic volume with $f_{\rm bov}=1$. All simulations commence at $z=127$, prior to this redshift we do not hazard a guess at the physical evolution and complex sociological structures that may have arisen in our space-borne herd. Instead, we allow the reader to imagine what may have transpired; it suffices to say there is a population of \mooons\ distributed as baryons would be in our own Universe. The two (12.5 cMpc)$^3$ volumes were allowed to run to $z=0$ to inexpensively probe the low redshift Universe, while the (50 cMpc)$^3$ volume was stopped at $z=2$ to probe the evolution of galaxies from the Epoch of Reionisation to Cosmic Noon.

We employ a modified version of the open source SPH With Inter-dependent Fine-grained Tasking \citep[SWIFT,][]{Schaller2018} code, with the SWIFT implementation of the EAGLE subgrid model \cite{Crain2015, Schaller2015, Schaye2015} presented in \cite{Borrow2022}. The bovine-based modifications to the code and model are detailed in Section \ref{sec:cow_subgrid}.


Instead of generating a stand-alone set of initial conditions (ICs) we choose to take advantage of the publically available EAGLE ICs packaged within SWIFT. Doing so enables direct comparison to the original EAGLE suite of simulations, and thus a bovine-free universe without additional computational resources. For details on the generation of these ICs we direct the reader to the EAGLE paper \cite{Schaye2015}.

To extract galaxies from our simulated \mooon+bovine-ejecta density field we employ Velociraptor \citep{Elahi2019}\footnote{Setting a Velociraptor on a field of bovines may sound concerning on the surface but we assuage any worries by reminding the reader these are simulated representations and a clustering algorithm. No bovines were harmed.}. Throughout this work we ensure robust measurements by only including galaxies with $M_\star> 10^8 {\rm M}_\odot$.

\section{\mooon\ Physics}
\label{sec:cow_subgrid}

There exist no thorough (or indeed cursory) studies of the density necessary to convert a cloud of cows into a star. Alas, due to this limitation of the literature, we must necessarily employ some assumptions about our primordial bovine herd (PBH).

\begin{itemize}
    \item On cosmological scales the PBH can be treated as a fluid.
    \item \mooons\ have a similar chemical composition to cows on Earth. Alas, yet again there is surprisingly little literature on the chemical composition of a cow, although there is a wealth of studies on the chemical composition of milk (a clear example of humanity's self-obsessed nature). Instead, we assume a chemical composition similar to that of humans shown in Table \ref{tab:abundance} which we propose is more than sufficient for our use.
    \item Unlike \cite{Roper2022}, we require \mooons\ to be compressible. Although they may seem to be incompressible on Earth, under star-forming conditions the authors are confident densities can be reached at which a cow is sufficiently squishy.
    \item Once the densities necessary for star formation have been achieved, \mooons\ will have degraded into a fluid capable of forming stars.
    \item \mooons\ are capable of maintaining a semi-consistent ratio between the number of \mooons\ and \mooon\ ``products''.
    \item Although \mooons\ are not spherical cows we do acknowledge that the spherical collapse of star-forming gas clouds does enable a certain number of ``spherical cow'' adjacent jokes to be applicable.
\end{itemize}

\begin{table}[h]
\centering
\caption{The assumed chemical composition of a \mooon, based on \citep{Chang2007}.}
\begin{tabular}{|l|c|}
\hline
Element & Proportion (by mass) \\
\hline
Oxygen & 65\% \\
Carbon & 18\% \\
Hydrogen & 10\% \\
Nitrogen & 3\% \\
Calcium & 1.5\% \\
Phosphorus & 1.2\% \\
Potassium & 0.2\% \\
Sulfur & 0.2\% \\
Chlorine & 0.2\% \\
Sodium & 0.1\% \\
Magnesium & 0.05\% \\
Iron & $<$ 0.05\% \\
Cobalt & $<$ 0.05\% \\
Copper & $<$ 0.05\% \\
Zinc & $<$ 0.05\% \\
Iodine & $<$ 0.05\% \\
Selenium & $<$ 0.01\% \\
\hline
\end{tabular}
\label{tab:abundance}
\end{table}

One might be inclined to wonder how the methane production utilised so successfully in \cite{Roper2022} might affect the evolution of \mooons. Indeed, the authors acknowledge that this is an important consideration. However, given the implied diet of our PBH, we cannot with certainty predict a \mooon's methane production rate\footnote{A \mooon\ producing methane at the same rate as Earth cattle \citep[$3.0 \times 10^{-6}$ kg $/$ s, e.g.][]{methRate1, methRate2, methRate3, methRate4}) would produce its own mass in methane in approximately 7 years.}, and under no circumstances are we calling for studies into this matter. Even so, \mooons\ will produce methane, faecal matter and other gases which must be considered, we denote these ``bovine-ejecta''. Luckily, since our simulations lack the resolution to resolve individual \mooons\ and elemental abundance must be conserved we can consider baryonic particles in our simulation as an amalgamation of \mooons\ and bovine-ejecta with constant macroscopic chemical properties. Below the resolution of the simulation, these two baryonic components are undergoing complex exchanges between the two states without any changes to the macroscopic properties of our baryonic simulation particles. We also note that below the resolution of the simulation, the bovine planetoids studied in detail in \cite{Roper2022} are extremely abundant and are analogous to brown dwarfs produced by failed star formation in our own Universe. Incidentally, the presence of such objects with a high fraction of organic material may offer the opportunity for the development of other lifeforms in the Universe.

It is highly likely that during periods with an overabundance of bovine-ejecta relative to the abundance of \mooons\ the star formation efficiency of a \mooon\ population could be enhanced, but we leave this consideration to a detailed future study. 

Thus the subgrid model is deceptively simple. \mooons\ are initialised with their chemical composition and distributed uniformly throughout the initial conditions to achieve the desired bovine fraction ($f_{\rm bov}$) while maintaining the primordial density field. Any normal baryons remain unchanged. The bulk of the model then remains unchanged with the same radiative cooling, star formation, stellar feedback, black hole seeding, and AGN feedback prescriptions used in the EAGLE model acting on the PBH, for a detailed description of the EAGLE model see \cite{Crain2015}, \cite{Schaller2015}, \cite{Schaye2015} and \cite{Borrow2022}. 

\section{Galaxy Evolution}
\label{sec:results}

With no prior studies into \mooon\ dominated cosmology, we have no intuition for what a galaxy would, or indeed should, look like when birthed from the \mooon+bovine-ejecta density field. In this section, we will take the \mooon\ by the horns and endeavour to elucidate this mystery. 

\subsection{Cosmic Star Formation Rate Density}

\begin{figure}[H]
    \centering
    \includegraphics[width=0.8\textwidth]{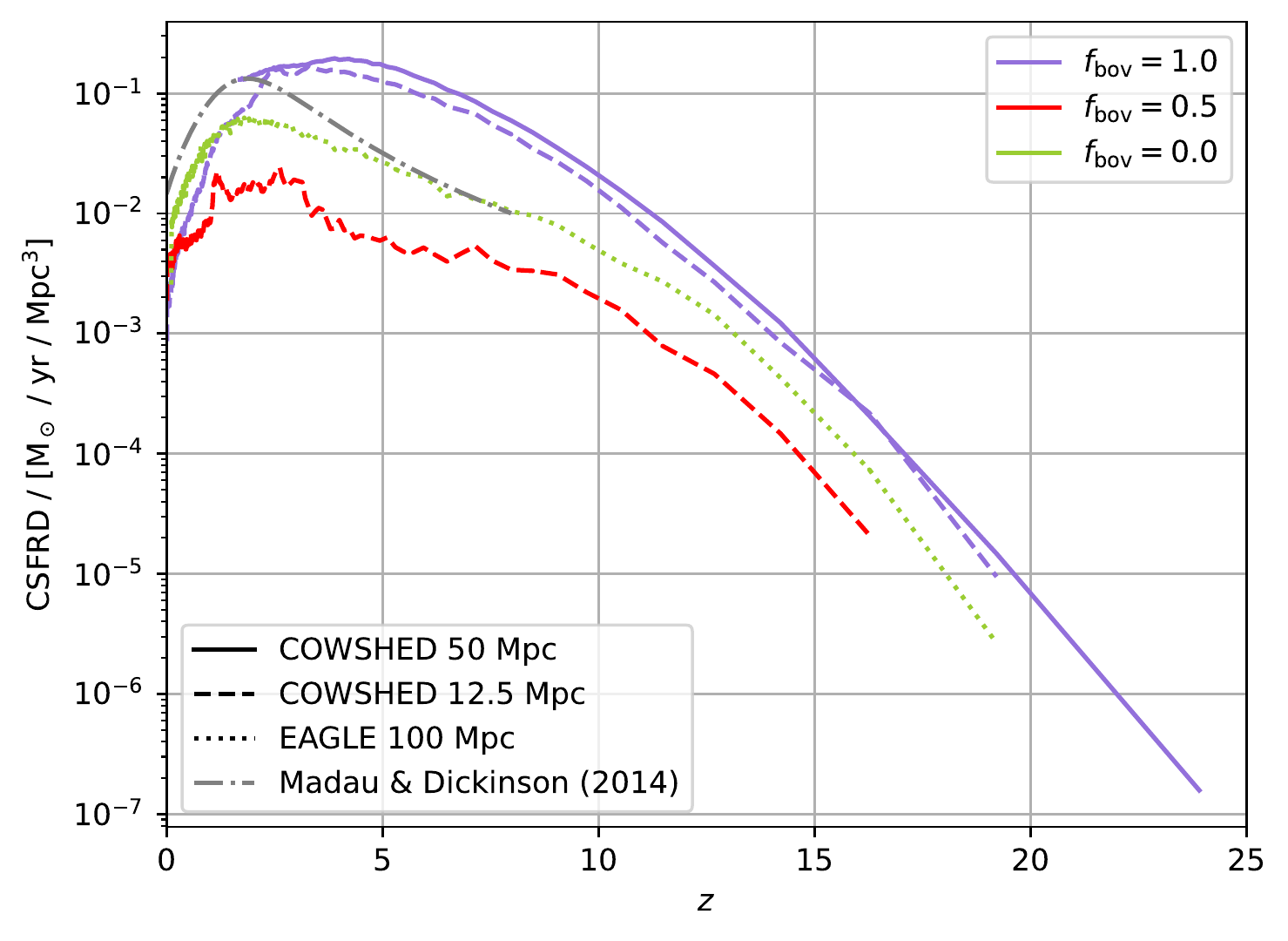}
    \caption{The cosmic star formation rate density (CSFRD) comparing a PBH-free cosmology to various bovine fractions. Each linestyle represents a particular volume or realisation of the CSFRD, while the colour dictates the simulated bovine fraction ($f_{\rm bov}$). We include the fiducial (100 cMpc)$^3$ EAGLE simulation \citep{Schaye2015} (green dotted line) and CSFRD from \cite{Madau2014} (grey dashed and dotted line) for comparison.}
    \label{fig:csfrd}
\end{figure}

We begin this illuminating investigation by presenting the global evolution of star formation in a variety of different cosmologies. In Figure \ref{fig:csfrd} we present the Cosmic Star Formation Rate Density (CSFRD) including comparisons to the CSFRD from the fiducial (100 cMpc)$^3$ EAGLE simulation and the observationally derived CSFRD from \cite{Madau2014}. 

Focusing on the pure-bovine cosmologies with $f_{\rm bov}=1$ we can clearly see an increase in global star formation for $z>2$ relative to both comparison curves. This increase should not be unexpected, the primordial chemical enrichment provided by our PBH allows for the effective cooling of the \mooon+bovine-ejecta density field enabling higher densities, and therefore star formation, significantly earlier than the \mooon-free universe. However, this increased star-forming potential can only last so long. We can see that in a \mooon\ dominated universe Cosmic Noon is shifted from $z=2$ to $z\sim4$. After which the cosmic star formation rate begins to drop, as evident in the (12.5 cMpc)$^3$ volume run to $z=0$.  

Interestingly, the enrichment provided by a PBH seems to induce an extreme fine-tuning problem to the universe in which they reside. We can see the universe with $f_{\rm bov}=0.5$ exhibits significantly suppressed star formation across cosmic time relative to its volume counterpart with $f_{\rm bov}=1$. This suppression is the result of the required balance between complex elements to aid cooling and star-forming matter. While one might assume the bovine-ejecta enrichment of the non-bovine gas should yield increased star formation efficiency, there is in reality a bovine fraction tipping point at which the increase in cooling efficiency, due to the complex elements present in the PBH, is balanced out by the reduced abundance of Hydrogen. The result is a fine-tuning problem which we defer to future April fool's days.

Evidently, not only are there numerous failed universes with ill-prepared \mooon\ fields at the time of recombination, but there are numerous successful \mooon\ universes with highly inefficient star formation. This inefficiency leads to universes with limited capacity for the formation of Earth-like planets containing cowsmologists capable of understanding their place in the bovine heavens. In all subsequent analyses, we will focus on the more enticing case of a \mooon\ cosmology with $f_{\rm bov}=1$.

\subsection{Galaxy Stellar Mass Function}

\begin{figure}[H]
    \centering
    \includegraphics[width=0.8\textwidth]{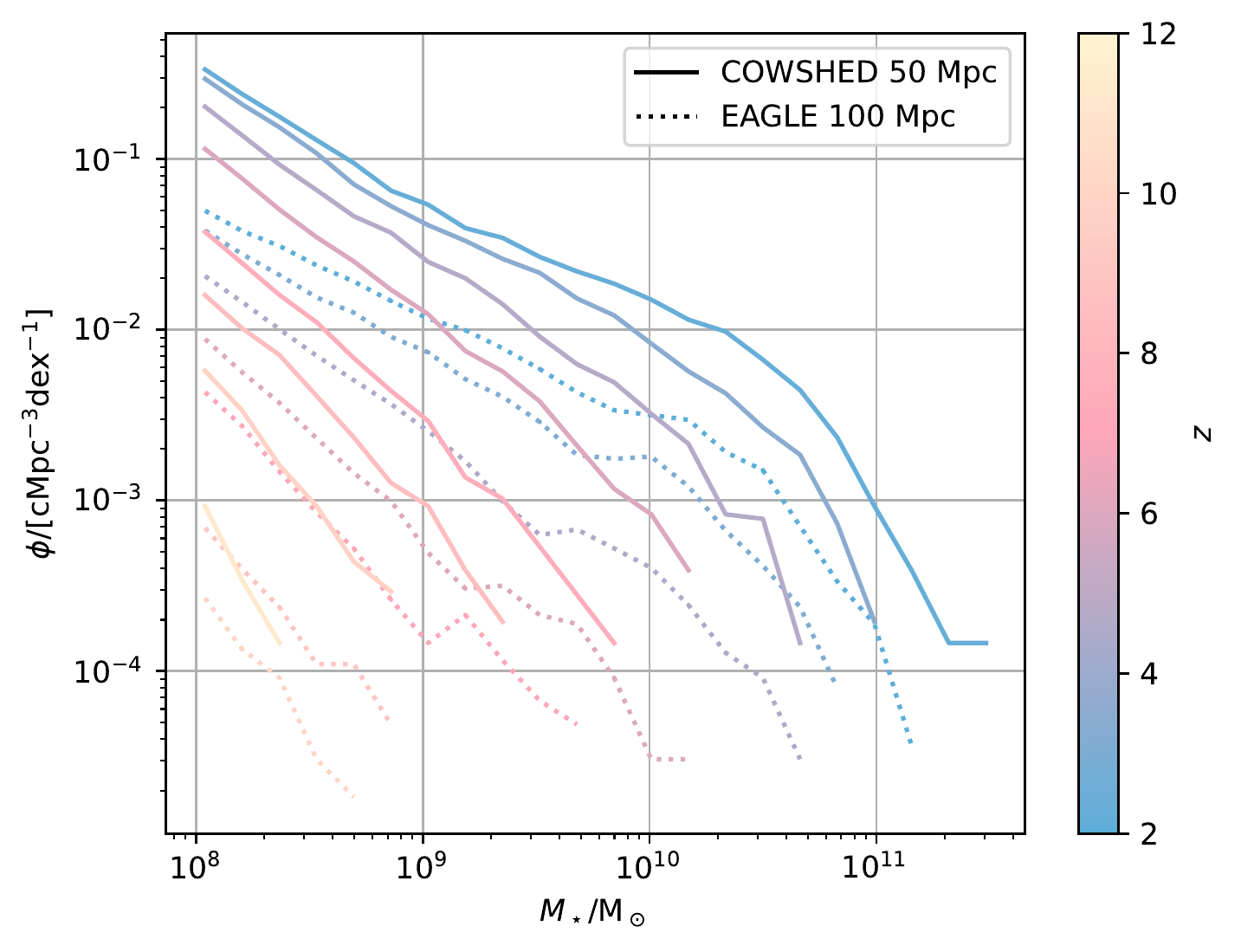}
    \caption{The Galaxy Stellar Mass Function at $z>2$ in the (50 cMpc)$^3$ \mooon\ dominated ($f_{\rm bov}=1$) simulation and the fiducial (100 cMpc)$^3$ EAGLE simulation \cite{Schaye2015}. Solid lines show the COWSHED GSMF while dotted lines show the EAGLE GSMF.}
    \label{fig:gsmf}
\end{figure}

Given the increased cosmic star formation rate in a \mooon\ dominated universe, it stands to reason that galaxies should begin forming earlier and thus result in significantly more massive stellar systems. Indeed, in Figure \ref{fig:gsmf} we present the Galaxy Stellar Mass Function (GSMF) in the \mooon\ dominated universe compared to the fiducial (100 cMpc)$^3$ EAGLE simulation and this postulate is plainly true. Across the whole presented redshift range the \mooon\ derived galaxies are consistently $\sim0.5-1$ dex more massive than those yielded by the EAGLE model at fixed redshift. 

\subsection{Observational Comparison}

To perform a more thorough investigation of the differences between \mooon\ derived galaxy populations and those present in our own Universe, we now present the Star Formation Rate Function. The Star Formation Rate Function (SFRF) can be readily derived from the observed UV luminosity function. We employ the conversion presented in \cite{Dayal2022} assuming a Chabrier IMF \citep{chabrier_galactic_2003},
\begin{equation}
\label{eq:sfrconv}
    \mathrm{SFR} = \kappa L_{\rm UV},
\end{equation}
where $\kappa=7.1 \times 10^{-29} [\mathrm{M}_\odot \ \mathrm{yr}^{-1} \ \mathrm{erg}^{-1} \ \mathrm{s} \ \mathrm{Hz}]$.

\begin{figure}[H]
    \centering
    \includegraphics[width=0.8\textwidth]{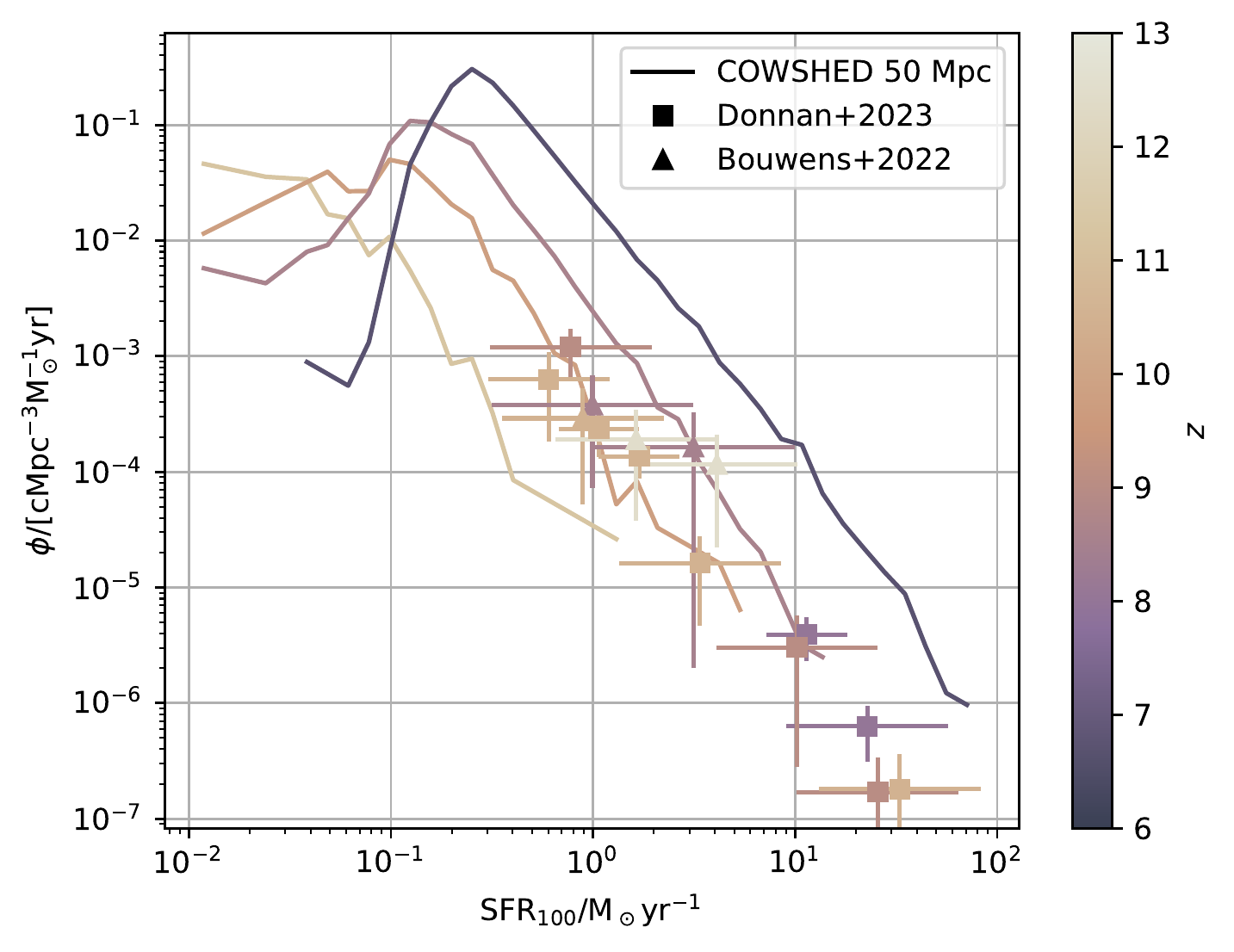}
    \caption{The Star Formation Rate Function (SFRF) at a range of redshifts for COWSHED and observations. The lines show the SFRF for the (50 cMpc)$^{3}$ COWSHED simulation. We include comparisons to early JWST observations from \cite{Bouwens2022} and \cite{Donnan2023} which have been converted from their presented UV luminosity function assuming a Chabrier IMF as per \cite{Dayal2022}.}
    \label{fig:sfrf}
\end{figure}

In Figure \ref{fig:sfrf} we present the SFRF derived from \cite{Bouwens2022} and \cite{Donnan2023} using Equation \ref{eq:sfrconv} compared to the galaxies nurtured by our PBH. We only include \mooon\ based galaxies with $M_\star> 10^8 {\rm M}_\odot$. Astoundingly, we can see a phenomenal agreement at $z\sim8-10$ between these high redshift observations and the galaxies present in a \mooon\ dominated universe. 

Much has been made of the challenge to $\Lambda$CDM the first JWST observations have posed. Given this fantastic agreement in terms of galaxy star formation rate, we can only conclude that a \mooon\ based cosmology should be added to the pile of theories competing to explain the discrepancies between high redshift observations and accepted theories. Of course, we have shown that the low redshift CSFRD does not agree between ours and a \mooon\ dominated Universe. Not to mention that galaxies are significantly more massive than those produced in the EAGLE model. However, there is a clear explanation. One that we have not included in the model but has a minimal effect at these early times when we find agreement. This explanation not only explains these discrepancies but also solves a long-standing mystery in cosmology: dark energy. 

We postulate that there could exist a mechanism, on the specifics of which we will not speculate, by which \mooons\ decay producing dark energy. In this conversion mechanism, a \mooon\ will decay producing Hydrogen, enabling the observed effects of reionisation, and dark energy solving two problems with one bovine (well, a universe full of bovines but that is just semantics). One enticing facet of this postulate is that the ubiquity of \mooons\ in a \mooon\ dominated cosmology removes any requirements for ``spooky action at a distance'' to explain dark energy.

\section{Conclusion}
\label{sec:conc}

We have presented the first study into galaxy formation and evolution in \mooon\ based cosmologies. We find that the introduction of \mooons\ can increase the cosmic star formation rate significantly resulting in more massive galaxies at fixed redshift relative to other models. We also find that a \mooon\ based cosmology gives rise to a new fine-tuning problem in the bovine fraction $f_{\rm bov}$, a manifestation of the balancing act between bovine enrichment of star-forming matter and the abundance of star-forming matter.

The initial goal of this work was merely to probe what such a universe may look like. However, we find that high redshift galaxies in our very own Universe agree shockingly well with the Star Formation Rate Function (SFRF) of \mooon\ reared galaxies in a universe with $f_{\rm bov}=1$. We thus conclude that our own Universe could be governed by \mooon\ cosmology. However, for this to be true \mooons\ must decay to yield the observed low redshift galaxy samples. We postulate that this decay is the mechanism by which dark energy comes into existence and begins to dominate the energy budget of the Universe.

This investigation has offered only a preliminary analysis of the consequences of a cosmology including \mooons, but we hope we have opened the gate to a lush new field of research possibilities.
Future work may be able to explore the internal structure of bovine galaxies, the mechanism for star formation from a PBH, and the evolution of cow-dependent lifeforms; further milking the enormous potential of the COWSHED project.

\backmatter





\bmhead{Acknowledgments}
We would like to thank everyone subjected to hearing about this work for putting up with the year-long discussion surrounding this project. The confused looks and chuckles maintained the drive to get this work completed. 

As with \cite{Roper2022}, we wish to acknowledge Jeff Bezos for his comments on space-based factories, which (for reasons that are but a distant memory) somehow led to
a lunchtime conversation derailing into the concepts presented in \cite{Roper2022}, which inevitably led to the work presented here.

We note that this work was conducted using minimal computational resources and low-priority queues to ensure it did not waste resources and get in the way of other's projects.

\bmhead{Data Availability}
All data is available upon request. The code used to run the simulations presented in this work is a fork of SWIFT and can be found \href{https://github.com/WillJRoper/swift-cows}{here}. The analysis scripts are similarly available \href{https://github.com/WillJRoper/cowshedII_analysis}{here}.

\subsubsection*{Software Citations}

This paper made use of the following software packages: SWIFT \citep{Schaller2018}, VELOCIRAPTOR \cite{Elahi2019}, MATPLOTLIB \cite{matplotlib}, SCIPY \cite{scipy}, NUMPY \cite{numpy}, UNYT \cite{unyt}, SWIFTSIMIO \cite{swiftsimio}, and SWIFTASCMAPS \cite{swiftascmaps}.










\begin{appendices}






\end{appendices}


\bibliography{cowshedII}


\end{document}